\documentstyle[12pt]{article}

\newcommand{\Section}[1]{\section{#1}\setcounter{equation}{0}}

	 \def\be{\begin{equation}}
	 \def\bea{\begin{eqnarray}}
	 
	 \def\o{\over}
	 \def\ee{\epsilon}

	 \def\ee{\end{equation}}
	 \def\eea{\end{eqnarray}}
	 \def\R{\rm {I\kern-.200em R}}
	 \def\C{\rm {I\kern-.520em C}}
\begin{document}

\vspace*{-2truecm}
\begin{tabbing}
  \` IPM-95-90
  \\
  \` August 1995
  \\
\end{tabbing}
\vskip 10mm
{\large {\bf
	  \centerline{
		    Expansion of Bubbles in Inflationary Universe
			} 
	 } 
 } 
\vskip 10 mm

\centerline{
	      Masoud Mohazzab\dag\ddag\footnote{ mohazzab@het.brown.edu.  Current
address: Box 1843, Physics Dept. Brown University, Providence, RI 02912, USA}
	     }
\centerline{ Mohammad M. Sheikh Jabbari\S
		}
\centerline{ and}

\centerline{Hadi Salehi\dag\footnote{salehi@netware2.ipm.ac.ir}
		}
\vskip 10 mm
{\it
  \centerline{\dag
	      Institute for Studies in Theoretical Physics and
	      Mathematics,
	     }
  \centerline{
	     P.O.Box  5746, Tehran 19395, Iran.
	     }
  \centerline{
		and}
  \centerline{ \ddag Physics Dept. Alzahra University
	      }
  \centerline{
	       Tehran 19834, Iran
	       }
  \centerline{ \S Physics Dept. Sharif University Of Technology
		}
   \centerline{ P.O. Box. 11365-9161, Tehran, Iran
		}
    } 

\vskip 10 mm

\begin{abstract}

We show that particle production during the expansion of bubbles of true vacuum
in the sea  of false vacuum is possible and calculate the resulting  rate.   As
a result the nucleated bubbles cannot expand due to the transfer
of false vacuum energy to the created particles inside the bubbles.  Therefore
all the inflationary models dealing with the nucleation and expansion of the
bubbles (including extended inflation) may not be viable.

\end{abstract}

\newpage

\Section{
	  Introduction
	  }

The idea of inflation (\cite{GUTH}, \cite{RHB}) solves many
 problems of old cosmology such as horizon problem.
 The main idea in inflationary models is the rapid
 expansion of the early universe ($e^{50}$ times the initial value) in a very
short time. (In fact inflationary models are not the only candidate that can
solve the horizon problem.  There is a recent claim that this problem can be
solved as a by-product of attributing a variable dimension to the universe
\cite{KMME}.)  The
original inflationary scenario  \cite{GUTH} was that the universe
started from a de Sitter space-time and bubbles formed randomly
in the sea of the de Sitter space-time through tunnelling of the inflaton
field.
The bubbles expand with the speed of light after
their nucleation. The  bubbles expand because the energy of the
 false vacuum is transferred to the kinetic energy of their wall \cite{COL}.
The bubbles, however, never percolate in the old inflationary
model (graceful exit problem).

Among  other surviving models of inflation is
Extended Inflation \cite{LS}.
In the extended inflationary model the problem of percolation of bubbles is
solved by substituting scalar-tensor
gravity theories (such as extended Branse-Dicke actions)
 instead of Einstein Hilbert action.  The bubbles nucleate
with a variable rate so that their
number will be enough to fill the false vacuum and the graceful exit problem
of the original inflationary model is solved.

  So far it has been assumed that the bubbles of true vacuum  can expand
  in the false
vacuum up to the time when they meet one another and mix into one.
This is the way the inflation is ended and
the universe ends up in true vacuum.

 Figuring out the exact space time in the interior of the bubble is a
 formidable task.  For the case when the space-time is homogenous but
anisotropic it has been shown that the anisotropy suppreses bubble nucleation
rate \cite{MM} but, so far, there is no discussion about the interior of the
bubbles.  For the case of isotropic models, it is usually assumed that the
space-time inside the
 bubble's wall is Minkowski while that of the outside is de Sitter
 or de Sitter like ( as is the case for extended inflation) \cite{LS}.

 An expanding  bubble causes an observer which was located in the space-time of
 false vacuum finds himself in the space-time
of true one after the bubble's wall expands and passes him.
 The very change of space-time should cause particle production and
therefore some part of the energy of the false vacuum  transfers
to them.  As a result the energy of the false vacuum
will be divided between two parts one part is the
kinetic energy of the bubble's wall and the other the produced particles.
If the transfer of energy to the created particles
is high enough (more than the kinetic energy of the
bubble's wall) the bubble won't expand and therefore it may stop
growing.  As a result the universe will be full
of bubbles some of them stationary and some other expanding slowly in
de Sitter space-time.  The problem to guess  the fate of these bubble
 is not known yet.  The production of particles during vacuum
tunnelling in Minkowski space-time has been widely studied \cite{RU},
\cite{KAN} and \cite{YTS},
however the quantum situation when bubbles expand is not known and is the
main subject of the present work.

 Here we investigate the quantum fluctuations of vacuum during the expansion
 of bubble in an homogenous and isotropic universe.
In what follows we will bring, after stating the idea of this work, the
calculations of the amount of particles created during the bubble expansion.
 Here we take the space-time
 inside and outside the bubbles as Minkowski and de Sitter respectively.
This rate is caculated in the approximation when the bubble is large enough to
attribute zero temperature to the inside
and also for large conformal time to be able to define the vacuum state.
Making these assumptions is important
in order to make our quantum field theory method valid. The exact evaluation
of the matrix elements needs the knowledge of the metric all over the
space-time.
The metod we use for matching the metrics inside and outside the bubble follows
\cite{KM} and will be published later \cite{MSS}.

An improved method of calculating the rate at the early times of bubble
nucleation will be given in \cite{MSS}.
Our calculation may also be generalized to domain wall and any other thin
 walls when they expand \cite{SW}.

\Section{
Particle production due to the expansion of bubbles }

Here we consider the evolution of bubbles with false or true vacuum inside.
The general
view of the problem is that due to the expansion of the bubbles
the space-time  changes with time.  As
a result we expect to have particle production during bubble expansion.  This
 has crucial effects on
the inflationary universe models involving bubble formation and expansion
e.g. old and extended
inflation.  In what follows we calculate the rate of energy loss by bubble
expansion.

Assume two different space-times, denoted by the metrics
$g_{\mu \nu}^{(1)}$ and $g_{\mu \nu}^{(2)}$, separated by a
thin wall.
For a massive
scalar fields in the background of each of these space-times, the general
Lagrangians
are

\be
 L^{(i)}={1 \o 2}[-g^{(i)}]^{1 \o 2}
 [{g^{(i)\mu \nu}} \phi^{(i)}_{,\mu} \phi^{(i)}_{,\nu} -
[m^{(i)^2} + \xi R^{(i)}] {\phi^{(i)2}}]
\ee
where $i$ is 1 or 2, $m^{(i)}$ is the mass of the field $\phi^{(i)}$,
$R^{(i)}$ is the scalar curvature of
the space-time $(i)$ and $\xi $ is a numerical factor.

The field equation for the above lagrangian is

\be \label{KG}
(\Box_x^{(i)} + m^{(i)2} + \xi R^{(i)}) \phi^{(i)}= 0
\ee

$\xi =0$ is the choice for the minimal coupling and
$\xi = {1\o 4} {{(n-2)} \o {(n-1)}}\equiv \xi(n)$, where n is the space-time
dimension,
is the conformal coupling when the action for massless  field
is invariant under conformal transformation.

The metric around the bubble's wall can be written as

\be
d\bar s^2= C^2(\bar \eta)( -d\bar \eta^2 + d\bar r^2 + \bar r^2(d\Omega^2))
\ee
\be
d s^2= -d \eta^2 + d r^2 +  r^2 d \Omega^2
\ee
where ${\bar s}$ referes to the metric outside the bubble and  for the de
Sitter spacetime we have $C(\eta)={1 \o {H \eta}} $ with $H$ a constant.
For small values of the conformal time $\bar \eta $ it is nontrivial to
have a vacuum state outside the bubble.  However assuming $\bar \eta$ to be
sufficiently large
we can have an adiabatic vacuum
\be
{lim}_{\bar \eta \rightarrow \infty} {\dot C(\bar \eta) \o {C(\bar\eta)}}
\rightarrow 0
\ee
Therefore a vacuum state can be defined asymptotically for outside region.

The state inside is not a real vacuum state, but rather a thermal state
\cite{YTS}, however it is Lorentz invariant \cite{YTS}.  Here we assume a large
bubble so that the temperature inside the bubble
be low enough to be able to assume a vacuum state inside.

Now we can decompose a field in each of these space times in terms of
creation and annihilation operators $a_k$ and $a_k^+$

\be
\phi^{(i)}(x)=\Sigma_k {[a_k^{(i)} u_k^{(i)}+ a_k^{(i)+}u_k^{(i)*}]}
\ee

The solution for $u_k^{(i)}$ is given by \cite{BD}
\be
u_k^{(1)}={ {1 \o {\sqrt {2 \pi}}}e^{i k \bar x} {\sqrt {\pi \bar \eta}
H_{\nu}^{(2)}{(k\bar \eta)} }}
\ee
where $H$ is a Hankel function with $\nu^2= {9 \o 4} - 12 (m^2 R^{-1} +\xi)$
and
\be
u_k^{(2)}= {1 \o {\sqrt {2 \pi}} } e^{i k x}
{e^{i \omega_k \eta} \o \sqrt {2\omega_k }}
\ee
where $(1)$ and $(2)$ refer to outside and inside  the bubble, respectively.

Now we write the solution over all space time as follows
\be \label {phi}
\phi = \phi^{(2)}\theta (R_b-r) + \phi^{(1)}\theta (r-R_b)
\ee
where $\phi^{(i)}$ is the quantum field inside (for $i=2$) or outside (for $
i=1$) of the bubble.

The coefficient $\beta$ in the Bogolubov transformation is then the inner
product of the mode solution

\be \label {beta}
\beta_{k'k}=-(u_{k'}^{(1)}, {u_k^{(2)}})=
i\int_{\eta_0} {u_{k'}^{(1)}(\bar x(x)) {\stackrel{\leftrightarrow}
{\partial_t}} {u_k^{(2)}}^*(x)} d^3x
\ee
 Note that ${\bar x}$ refers to the coordinate outside the bubble.

 In fact the arguments in the inner product  are in different coordinates and
with the methods (e.g. \cite{KM}) of matching of the metrics the relation
between the two coordinate can be found \cite{MSS}.

The total number of particles per mode will be
\be \label{N}
N_k=\Sigma_{k'}\vert \beta_{k'k} \vert^2
\ee

The local observer previously in  de Sitter space-time, now finds itself in
Minkowski, where he sees particle production by the rate predicted by
(\ref{N}).

 From (\ref {beta} ) it is clear that when $u_k^{(1)}=u_k^{(2)}$, the
coefficient is zero.
 Therefore in our approximation i.e. $\bar\eta \rightarrow \infty$, the
coefficient, $\beta$, is very small.  However as we go back in time the
discripancy between  $u_k^{(1)}$ and $u_k^{(2)}$ gets larger and therefore the
rate of
particle production inside the bubble gets amplified.  As a result the bubble
may not have expanded to what we have considered as an approximation.

\Section {Conclusion}

The particle production rate due to the sudden change of vacuum in time
caused by the expansion of bubbles that separate two different space times has
been calculated.
Some part of the energy of the false vacuum then transfers to create particles,
rather than accelerating the wall and therefore
 the bubble will slow down expanding.  Therefore the inflationary models
concerning the production and expansion of bubbles may have serious graceful
exit problem, that the de Sitter universe will be
filled with static, or slowly growing, bubbles of true vacuum inside.  In fact
the state inside  these static bubble is a
hot thermal one partly induced by the particle production.

Here we made the assumption of big bubble and large conformal time in order to
be able to define vacuum states inside and outside of the bubble.
 With these assumptions we observed that the expansion of the wall of the
 bubble causes particle creation although very small.  However, this result
suggests that if we go back in
 time the quantum effect will be amplified and presumably the bubble could not
 have expanded to what we have considered as an approximation, that is
 the bubbles would have  stopped growing at the early times after their
nucleation.
 In fact in the recent work \cite{La}, it has been shown that the particle
 production is enhanced in a thermal state, which corresponds to
  small bubbles or at small conformal time in our model, and justifies our
suggestion.

  So this work shows that the
 inflationary models concerning the creation and expansion of bubbles
   have the difficulty explaning the graceful exit problem.

 The method explained in this work may also be applied to many other
situations, when domain wall or bubbles are concerned ( such as \cite{SW}).

 \section*{Acknowledgments}
We are grateful to Robert H. Brandenberger, Reza Mansouri and Martin Rainer for
helpful discussions.
M. M.  thanks for hospitality at the Projektgruppe
Kosmologie of Universit\"at Potsdam and their invitation to
present this work and the discussions.

\end{document}